\documentclass[a4paper, 12 pt]{article}

\usepackage{latexsym}
\usepackage{tikz}
\usetikzlibrary{decorations.pathmorphing} 
 \usetikzlibrary{decorations.pathmorphing,calc} 
 \usepackage{amsmath}
 \usepackage{graphicx}
 \usepackage{graphicx,wrapfig} 
 \usepackage{geometry}
 \usepackage{latexsym}
\usepackage{amsthm}
 \usepackage[titletoc,toc,title]{appendix}
\title{Geometric Backreaction  of Modified Quantum Vacua and Diffeomorphrisim Covarience }
\author{ Salwa Al Saleh }  \date{\textit{Department of Physics and Astronomey, College of Science, King Saud University. \footnotemark [1]}} 
\begin{document}
	\maketitle
	\renewcommand{\thefootnote}{\fnsymbol{footnote}}
	\footnotetext[1] { Email salwams@ksu.edu.sa}
\begin{abstract}
In this paper we have shown that squeezed modified quantum vacua have an effect on the background geometry by solving the semi-classical Einstein Field Equations in modified vacuum. The resultant geometry is similar to (anti) de Sitter spacetime. This geometry could explain the change of causal structure - speed of light- in such vacua without violating diffeomorphism covariance or causality.The superluminal propagation of photons in Casimir vacuum is deduced from the effective electromagnetic action in the resultant curved geometry. Singling between different vacua is shown not to violate causality as well when the geometric effect on the null rays is considered, causing a refraction of those rays when travelling between unbounded and modified vacua.
\end{abstract}
\section{Introduction}
Since the 50's light propagation in modified quantum vacua has been studied extensively. Several factors can affect the vacuum, including boundary conditions - Casimir vacuum- \cite{casimir} resulting a spacelike propagation of light rays knows as \emph{Scharnhorst effect} \cite{sharnhorstI,barton1993qed}. Heat bath can also modify the vacuum and result a variant speed of light depending on its temperature \cite{THERMAL,tarrach1983thermal}. Also the presence of external electric and magmatic fields would produce the same effect \cite{Latorre}. Most agree that this is a violation of  equivalence principle not causality, and those superluminal photons are evidence that diffeomorphism covariance may not be exact\cite{QEDgrav,shore1996}. However, this quantum nature of vacuum seems to explain why that speed of light is finite, as any contribution of 1-loop or 2-loop Feynman diagrams affects the Maxwell's action in modified vacua\cite{Barton1}. But it also rises the question about the exactness of diffeomorphism  covariance. This is a usual problem one faces when dealing with quantum fields in curved spacetime \cite{QFToncurvedspacetime}. Despite the tininess of this effect for many quantum electrodynamics (QED) or general relativity (GR) calculations\cite{impossibility}. It is a fundamental aspect of quantum field theory on curved background and even for a quantum theory of gravity. We propose that diffeomorphism covariance is exact and not broken by vacuum modification.  We still can make light always moves by null geodesics, but in different geometries. Causality is saved as well, even if singling occurs between observers living in different quantum vacua. 
\section{ Quantum Field theory on Modified Vacua }
\begin{figure} [h!]
	\label{fyenmandiag}
	\caption{An example of tow-loop Feynman diagrams having no-trivial contributions to Maxwell's action in modified vacua.}
	\includegraphics[scale= 0.75]{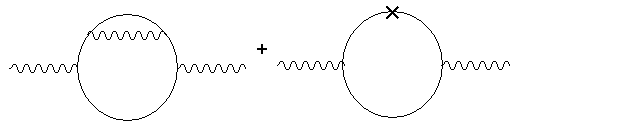}
	
\end{figure}

Consider an arbitrary scalar field $ \phi $. that satisfy the Klein-Gordon equation \\
\begin{equation}
\Box \phi =0 .
\label{eq KG}
\end{equation} 
This field satisfies the Green's function relations: 

\begin{subequations}
	\begin{alignat}{2}
	\langle0|[\phi (q,q')]|0\rangle &= iG(q,q'),\\
	\langle0|\{\phi (q,q')\}|0\rangle& = iG^0(q,q'),
	\end{alignat}
\end{subequations}
where $G$ is Pauli-Jordan function and $ G^0$ is known as Hadamard's elementary function.
We can express the field as a sum of the family of solutions in modes $ u_\omega= Ae^{i(kx-\omega t)} $:
\begin{equation}
\label{modsolution}
\phi= \sum_{\omega=0}^{\infty} a_\omega u_\omega + a^\dagger_\omega u^*_\omega .
\end{equation} 
However, if the field lies in a bounded vacuum, like Casimir vacuum, the modes ought to have frequencies larger than the cut-off frequency $\omega_0$. Since the particle creation is limited to frequencies larger than $ \frac{2\pi}{L} $ where $L$ is the distance between Casimir plates or the diameter of a spherically symmetric boundary condition. \cite{casimir}. However, there are other ways to modify the vacuum besides boundary conditions \cite{modify,sharnhorstI} - for example, thermal bath- \cite{THERMAL}. 
In a modified vacua, we should rewrite the field $ \phi $ in \eqref{modsolution} to become:
\begin{equation}
\label{modsolutionmodified}
\phi= \sum_{\omega> \omega_0}^{\infty} a_\omega u_\omega + a^\dagger_\omega u^*_\omega.
\end{equation} 
Surely this formulation can be generalized to massive fields and fields with arbitrary spin, in straightforward manner. When there is no excitation of the field, we consider the vacuum state. The unbounded vacuum is expected to have a non-vanishing energy term $ E= \sum \omega $. In this work however, we shall use the normally ordered Hamiltonian for the unbounded vacuum, getting rid of the divergent vacuum energy term. Implying that modifying a patch of the vacuum will perturb its energy. For example the energy perturbation contribution from  fluctuations of our scalar field can be calculated from \cite{jaffe2005casimir}
\begin{equation}
\label{peturbationG}
\langle \hat{ H}\rangle^{(1)} =  2\pi ^{-1} \Im \int d\omega \omega Tr \int d^nx\tilde{G} - \tilde{G}^0.
\end{equation}
Where  $ \tilde{G}$ and $\tilde{G}^0$ are the perturbed Pauli-Jordan/Schwinger green's function  and Hadmard elementary function, respectively. 
The energy perturbation can be calculated by zeta function regulation, for Casimir vacuum it can be seen in the literature \cite{casimir}:  
\begin{equation}
\label{ vacuumenergy}
\langle \colon \hat H ^{(1)\colon}\rangle =- \frac{\pi ^2}{720 L^4}.
\end{equation} 
Modified vacua do not just have a different zero-point energy, but also affect the propagation of photons or other massless particles within them. This effect has been studies extensively since the 1950's \cite{modify}. Scharnhorst and Barton \cite{sharnhorstI,Barton1,barton1993qed} had studied propagation of light between Casimir plates and concluded a refractive index less than unity. Implying a phase and group velocity of photons larger than 1.  The explanation behind that is the two-loop interaction between the photons and background fields- like Dirac field- (Figure\ref{fyenmandiag}). Will have a non-trivial contribution to Maxwell's action in modified vacua. The refractive index of a modified QED vacua can also be calculated from non linear QED as in \cite{liberati2001scharnhorst}. 

The refractive index for a Casimir is therefore given by :
\begin{equation}
\label{eqrefractive}
n=  1- \frac{11 \pi ^2 \alpha^2}{8100 \alpha_G^2 L^4 },
\end{equation}
where $ \alpha$ is the fine structure constant and $\alpha_G^2$ is the mass of the electron squared written in Plankian units (the gravitational coupling constant).The photons only interact with charged Dirac and gravitational fields. Using relation \eqref{eqrefractive} we can calculate the speed of light in Casimir vacuum to be:
\begin{equation}
\label{speedoflight}
c=1 - \beta_1,
\end{equation} 
( where $ \beta=- \frac{11 \pi ^2 \alpha^2}{8100 \alpha_G^2 L^4 }$ ).\par .
This is only an example of modified vacuum. Heat bath is shown to have an effect on the speed of light, the formula is given by \cite{tarrach1983thermal}
\begin{equation}
\label{speedoflightThermal}
c=1 - \frac{106}{2835} \frac{\alpha^2 \pi^2 T ^4 }{\alpha_G^2}.
\end{equation} 
Here, the speed of light is reduced by the thermal bath, same effects can be calculated in the presence of external fields, for instance. In all above examples the speed of light is either less or more than 1. An elegant  unified formula for the speed of light is given in terms of vacuum (fluctuation) density $ \rho_{vac} $ \cite{Latorre}
\begin{equation}
\label{unified}
c=1 - \frac{44}{135} \frac{\alpha^2 }{\alpha_G^2} \rho_{vac} ,
\end{equation} 
here $\beta =   \frac{44}{135} \frac{\alpha^2 }{\alpha_G^2} \rho_{vac} $ 
This beautiful formal applies to any modified vacua, hence the vacuum density might be positive - in case of thermal bath or external $E$ or $B$ fields-, or negative - for Casimir ( squeezed) vacuum.
Equation \eqref{unified} might appear to break the exactness of Lorentz symmetry. We shall demonstrate that a more general symmetry holds however, the diffeomorphism symmetry. Because of the geometric back-reaction.
 \section{Modified Vacua and Spacetime Curvature}
 The starting point of our analysis is to reformulate the quantum fields we discussed in the previous section in curved background. Basically, the same expansion of $ \phi $ in \eqref{modsolution} can be made in curved background. But care must be taken that  expansion \eqref{modsolution} is not unique for all frames \cite{QFToncurvedspacetime}. 
 Since the stress-energy tensor for the unbounded vacua is normally ordered, the vacuum expectation value of the stress-energy tensor operator is  related to the modified vacuum density $\langle \hat {T}_{\mu\nu}\rangle = \rho_{vac} g_{\mu\nu}$ Here we follow the usual treatment of vacuum as a perfect fluid with density $ \rho_{vac}$ \cite{curvedI}. The semiclassical Einstein field equations (EFE's) are therefore written as :
  \begin{equation}
  \label{EFE} 
  R_{\mu\nu}+ \frac{1}{2} \Big( R+ b \Big) g_{\mu\nu}=0 ,
  \end{equation} 
setting $ 2 \kappa \rho_{vac} = b$.\par
For Casimir vacuum $b$ is given explicitly from the expectation value of the normally-ordered stress-energy tensor of the quantum field described above:
\begin{equation}
b= 2\kappa \langle \colon \hat{T}_{00}\colon\rangle= 16 \pi \left( \frac{-\pi^2}{ 90 L^4}\right) 
\end{equation}
 These are EFE's in modified vacuum having . Surely to attempt solving \eqref{EFE} we need to specify the boundary conditions and symmetries.of the modified vacuum metric $ \tilde{ g}_{\mu\nu} $. However, care should be taken here, the factor $ b$ is not usually independent from the boundary conditions. For example, for Casimir vacuum the boundary conditions will change $b$ as itself has risen from these boundary conditions. Hence, to demonstrate the general principle it is enough to observe \eqref{EFE} as it stands without attempting to solve it first. The factor $b$ plays a r\^{o}le similar to the cosmological constant, regarding its effect on the metric.
 The modified metric corresponds to a negative ( positive) intrinsic curvature - depending on the value of $\rho_{vac}$. Therefore, modifying the vacuum is conformal transformation to the metric, viz, modified vacuum is a conformal vacuum, as we shall demonstrate later.
 
 To illustrate this result we shall take some examples of solutions of EFE in modified vacua. We begin with a non-relativistic mass density satisfying Poisson  equation and  $ S^2 $ symmetry :
 \begin{equation}
 \label{poisson}
 \text{Lap}( \Phi)  = 4 \pi \rho_{matter}.
 \end{equation}
 To solve Einstein field equations for this matter, we use linear perturbation theory assuming that the metric can be written as :
 \begin{equation}
 \label{peturbation}
 g_{\mu\nu} = \tilde{g}_{\mu\nu} + h_{\mu\nu},
 \end{equation}
 where $\tilde{g}_{\mu\nu}$ is the modified vacuum metric that solves \ref{EFE} and  $h_{\mu\nu}$ is the normal, gauge-invariant metric perturbation. We first consider the trace-reversed perturbation:
 \begin{equation}
 \begin{split}
 \bar{h}_{\mu\nu}&= h_{\mu\nu}-\frac{h}{2}\tilde{g}_{\mu\nu}, \\
 \Rightarrow& \bar{h}_{00} = \Phi .
 \end{split}
 \end{equation}
 Where $h$ is the contracted perturbation metric. Now we can easily find from the spherical symmetry the terms of the metric perturbation:
 \begin{subequations}
 	\[
 	h_{\mu\nu}= 
 	\begin{cases}
 	-2 \Phi -\frac{1}{2} bf(\textbf{x},t),& \text{when } \mu=\nu;\\
 	0,              & \text{when }  \mu \neq \nu.
 	\end{cases}
 	\]
 \end{subequations}
 The term $\frac{1}{2} bf(\textbf{x},t) $  came from modifying the vacuum, and the function $f$ depends on the solution of \eqref{EFE} , the Killing vectors fields of the modified vacuum metric. We write the solution to this perturbation as a modified Schwarzchild solution. Knowing $ M = \int d^3x \rho_{matter} $. 
 \begin{equation}
 \label{schwarzchild}
 ds^2=- \Big(1-\frac{2M}{r} +bf(\textbf{x},t)\Big) dt^2 +\Big(1-\frac{2M}{r} + bf(\textbf{x},t) \Big)^{-1}dr^2 + r^2d\Omega^2_2 .
 \end{equation}
 Thus, vacuum modification could either reduce or increase Schwarzchild radius depending on the value of vacuum density discussed earlier as a result of changing the background geometry, as if distances are changed in this transformation- apparent from the change of Schwarzchild radius present in \eqref{schwarzchild}-. \\ We clearly observe the. It will become much clearer as we discuss the propagation of photons in modified vacua. Now, we attempt to solve \eqref{EFE} for a maximally-symmetric vacuum satisfying $O(1,4)/O(1,3)$ symmetry. The solution is known as lambdavacuum solution: 
 \begin{equation} 
 \label{symmetrysolution}
 ds^2= \frac{3}{b\eta^2}\left(-d\eta^2+\sum^3_{i=1}dx_i^2\right) ,
 \end{equation}
where $\eta=e^{-\sqrt{\frac{b}{3}}t}$ the conformal time.
 This solution is valid inside the boundaries of modified vacua. Provided that $b$ is constant within the boundary, viz independent of spatial direction. We can see that This solution is clearly a conformal  transformation of the Minkowski metric $ \eta_{\mu\nu}$ and adds a global curvature with characteristic length $\frac{3}{b} $.
 For negative $b$ the above metric is written as anti de Sitter metric, like for example :
 \begin{equation}
 ds^2= \frac{3}{bx_1^2}\left(-dt^2+dx_1^2+\sum^3_{i=2}dx_i^2\right).
 \label{ads}
 \end{equation}
 The key elements for this solution are the curvature tensors and scalar, they are independent of the coordinates chosen to represent the solution. 
 Riemann and Ricci tensors , and Ricci scalar are written ( respectively) as:

 \begin{subequations}
 
\begin{gather}
 \tilde{R}_{\nu\alpha\beta}=-\frac{1}{\alpha^2}(g_{\mu\alpha}g_{\nu\beta}-g_{\mu \beta}g_{\nu\alpha}), \\
   \tilde{R}_{\mu\nu} = -\frac{3}{\alpha^2} g_{\mu\nu},\\
\tilde{R}=-\frac{12)}{\alpha^2} .
  \end{gather}

 \end{subequations}
Here we use the variable $ \alpha$ (radius of AdS) which is given by:
\begin{equation*}
b= - \frac{3}{\alpha^2} ,
\end{equation*}
notice that if $b<0$ the curvature is negative and the converse is true for $b>0$, and the Einstein tensor is found to be:
\begin{equation}
\tilde{G}_{\mu \nu} = -b \left( \begin{matrix} -1&0&0&0\\0&1&0&0\\0&0&1&0\\0&0&0&1\end{matrix} \right) .
\end{equation}
 This solution alone is not sufficient to specify the light rays. We need to consider quantum corrections to it to model light propagation in modified vacua. We can use the metrics \eqref{symmetrysolution} or \eqref{ads} to calculate the magnitude of a spacelike vector $ X^\mu$ finding shorter  or longer magnitude than its magnitude with respect to Minkowski metric depending on the sign of $b$. This shows that diffeomorphism covariance holds for modified vacua, as the speed of light did not actually change but the geometry ( as if the light has shorter distance to travel).  However, to formulate this more rigorously, one needs to study the effective metric for photon propagation in curved spacetime.
 
 \section{Photon Propagation in Modified Vacua and Effective Action in Curved spacetime}
 When trying to study the propagation of photons in curved spacetime, using the effective metric as it was shown by \cite{QEDgrav} that gravitational "tidal"effects also affects the speed of light, some interesting spacetimes had this effects like Schwarzchild \cite{QEDgrav} Kerr \cite{kerr} and Reissner-Nordstr\"{o}m  \cite{chargedBH} .   \\  Hence, using the effective metric with spacetime described in \eqref{symmetrysolution} should result photon propagation with speed given in \ref{unified}. Thus spacetime geometry change corresponds to all cases of modifying the vacuum. We start by writing the effective metric for QED in the spacetime solving equation \eqref{EFE}. 
  \begin{equation}
  \label{effectiveaction}
  \Gamma= \int d^4x \sqrt{-\tilde{g}}\Bigg[- \frac{1}{4} F_{\mu\nu}F^{\mu\nu}+\frac{1}{\sqrt{\alpha_G}}\Big( q \tilde{R}F_{\mu\nu}F^{\mu\nu}+p\tilde{R}_{\mu\nu}F_{\mu\rho}F^{\nu}_{\rho} + s\tilde{R}_{\mu\nu \rho \sigma} F^{\mu\nu}F^{\rho \sigma} \Big)\Bigg].
  \end{equation}
  Where $ q=\frac{ -1}{144} $, $p=\frac{13\alpha}{360 \pi} $ and $ s=\frac{ -1\alpha}{360 \pi} $. For photon propagation the Faraday tensor is written in terms of waves $ F_{\mu\nu} = A_{\mu\nu} e ^{i\theta} $ with phase $ \partial_\alpha \theta = k_{\alpha}$ and amplitude $ A_{\mu\nu} $ satisfying the Bianchi identity $ k_\lambda A_{\mu\nu} +  k_\nu A_{\mu\lambda}+ k_\mu A_{\lambda\nu}=0$ The photons should obey the null geodesic equation:
  \begin{equation}
  \label{geodesic}
  k^\mu k^\nu_{;\mu} =0 .
  \end{equation}
  However, this does not hold when we derive the propagation of photons from \eqref{effectiveaction}. We see that the geodesic equation becomes:
  \begin{equation}
  \label{equationofmotion }
  k^2 = -\frac{2p}{\alpha_G} \tilde{R}_{ij} k^i k^j +\frac{8s}{\alpha_G} \tilde{R}_{ijlm}k^i k^j n^l n^m,
  \end{equation}
  where $n^k$ is the polarization 3-vector. Since the modified vacuum space is not flat, the RHS is not zero. Using the values of curvature tensors and scalars written earlier, we can write the general formula explicitly:
   \begin{equation}
   \label{generalformaula}
   - \frac{64 \alpha^2}{45 \alpha_G }\rho_{vac} \tilde{g}_{ij} k^i k^j + \frac{-128 \alpha^2}{1080 \alpha_G}\rho_{vac} (\tilde{g}_{ij}\tilde{g}_{lm}-\tilde{g}_{il}\tilde{g}_{jm})k^i k^j n^l n^m.
   \end{equation}
This formula allows us to retrieve the expression for the (group) velocity of light in modified vacua that appears in \eqref{unified} easily. Also allows to generalise the results in \cite{QEDgrav,kerr,chargedBH}. Since the metric tensor  that appears in \eqref{generalformaula} could be for any spacetime. \\ Since the effect is geometric, we can always locally set the metric to be the flat metric. Hence, we still can say that light travels in null geodesics in this vacuum.
We should emphasise on an important point, we assume maximally symmetric/ flat space, where polarisation of propagating photons does not matter ( this is evident from the product of the polarisation vectors $n^{i}$). The case of parallel plates in Scharnhorst effect, we wont be having all the Killing vector fields of the metric in \eqref{symmetrysolution}, rather the Killing vector field would be only the direction perpendicular to the plates $ X_ \perp $.
 
   \section{Causality between Different Types of Vacua}
    It comes as a fundamental question how observers in different vacua could signal each other, preserving diffeomorphism and causality. 
       \begin{figure}[h!]
       	\label{causality}
       	\includegraphics[scale= 0.75]{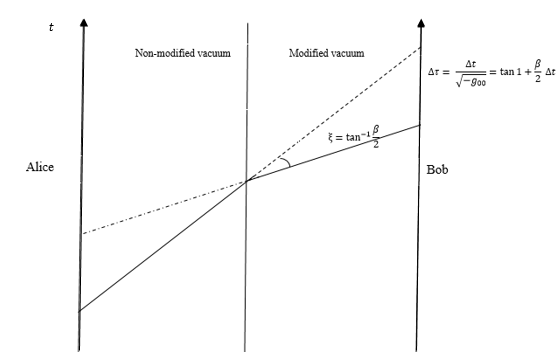}
       	       	\caption{Alice is in unbounded vacuum signalling Bob in a modified vacuum with spacelike separation ( with lower density).The null ray is refracted preserving causality. Bob will detect the signal as if It was sent later than the proper time of Alice had sent it.  }
       \end{figure}
    Specially as heat bath affects the state of vacuum. Without a law to let signalling occurs between different vacua. Quantum field theory is in real conflict with exactness of diffeomorphism. But the solutions come in a very simple manner: refraction. We know already that null rays travel always with a constant angle  ( $ \frac{\pi}{4}$ ) parallel to the null generators. This angle changes when the vacuum changes, it is a direct result of the geometry change . If light rays ought to travel between different vacua, it will face refraction. This is a result of wave properties of light, or naively from Snell's law ( but in spacetime expressed in terms of null coordinates).  If two observers in different types of vacua with a spacelike \footnote{It could be timelike saparation. However, that would be a different scenario as for example signals from the early universe this will not be discussed here} separation $ X $ try to signal each other , applying Snell's law we conclude first that they will only see light rays travelling at the speed of light in their own vacuum. However, each receptor will think that the signal is sent at a different time. In other words, the two observers do not have the same proper time, this is evident from metric variation due to modifying the vacuum.  $ t'= \frac{t}{\sqrt{-g_{00}}}$ . For example a receiver in a modified vacuum will preserve a signal from unbounded vacuum sent at time $t$ from the sender's perspective as if it was sent at $ t'= \frac{t}{\sqrt{g_{00}}}$ that is a little bit in the future. And vice versa when the receiver is in the unbounded vacuum, this will certainly preserve causality. Since both observers will not be able to send superluminal signals in their with respect to their reference frame. The refraction occurs as the shift between geometries is not smooth, it happens rapidly at the boundary.

     \section{Conclusion}
   The quantum vacuum puts the exactness of diffeomorphism covariance of general relativity into a threat. Since perturbing the quantum vacuum will have an effect on propagation of light in an apparent contraction with diffeomorphism covariance. Nevertheless, when the background back-reaction is taken into account, this contraction is resolved. We have shown that the background of non-trivial vacua admits geometry similar to (anti) de Sitter spacetime. If the non-trivial vacuum had a negative density, the background will be negatively curved ( hyperbolic), hence, if immersed into a flat spacetime distances between points will decrease compared to what is measured in flat background. This explains the apparent increased speed of light in such vacua. We may also conclude that the quantum nature of vacuum is important for stabilising the underlying geometry. Imagine what a severe perturbation on the vacuum fluctuations can do to the underlying geometry! As such perturbation will correspond to a large absolute value on $b$, making the spacetime extremely curved and unstable. This treatment of modified vacua is very important in the early universe as we are intending to show in future work.  
\section*{Aknowledgements}
 Warm regards to Prof. Nabil Bennessib for his generous technical help in writing this paper, and revising it. 
    
\bibliographystyle{plain}
\bibliography{ref}
\end{document}